\documentstyle[prb,aps,epsf,eqsecnum,graphicx]{revtex}

\newcommand{\be}{\begin{equation}}
\newcommand{\ee}{\end{equation}}
\newcommand{\bea}{\begin{eqnarray}}
\newcommand{\eea}{\end{eqnarray}}
\newcommand{\bd}{\begin{displaymath}}
\newcommand{\ed}{\end{displaymath}}
\newcommand{\bi}{\begin{itemize}}
\newcommand{\ei}{\end{itemize}}
\newcommand{\bc}{\begin{center}}
\newcommand{\ec}{\end{center}}
\newcommand{\bfl}{\begin{flushleft}}
\newcommand{\efl}{\end{flushleft}}
\newcommand{\bfr}{\begin{flushright}}
\newcommand{\efr}{\end{flushright}}
\newcommand{\f}{\frac}

\def\bk{{\bf k}} \def\bq{{\bf q}}

\def\ra{\rightarrow}
\def\6{\partial} \def\a{\alpha} \def\b{\beta}
 \def\d{\delta} \def\ve{\varepsilon}

\def\o{\omega} \def\G{\Gamma}

\def\={\!\!\!&=&\!\!\!}
\def\+{\!\!\!&&\!\!\!+~}
\def\-{\!\!\!&&\!\!\!-~}

\begin{document}
\draft
\renewcommand{\theequation}{\arabic{equation}}
\twocolumn[\hsize\textwidth\columnwidth\hsize\csname  
@twocolumnfalse\endcsname
\author{I. Tifrea$^{1,2}$, I. Grosu$^{2}$ and M. Crisan$^{1,2,3}$}
\address{$^{1}$Dipartimento di Matematica e Fisica, Sezione INFM, 
Universit\`{a} di Camerino, 62032 Camerino, Italy\\
$^2$Department of Theoretical Physics, University of Cluj, 3400 Cluj, Romania\\
$^3$``Abdus Salam'' International Center for Theoretical Physics, 
34014 Trieste, Italy}
\date{January 31, 2000}

\title{Magnetic instability of a two-dimensional Anderson non-Fermi Liquid}

\maketitle
\hspace*{-0.25ex}

\begin{abstract}
We show that in the Anderson model for a two-dimensional non-Fermi liquid a 
magnetic instability can lead to the itinerant electron ferromagnetism. The 
critical temperature and the susceptibility of the paramagnetic phase have 
been analytically calculated. The usual Fermi behaviour is re-obtained taking 
the anomalous exponent to be zero.
\pacs{PACS numbers:}
\hspace*{-0.25ex}

\end{abstract}]

\narrowtext

Since the discovery of high-$T_c$ superconductors there has been a great 
interest in the study of the non-Fermi liquid model in two dimensions (2D). 
The model proposed by Anderson \cite{1,2} has a phenomenological character 
because the form of the Green's function has been taken from the 
one-dimensional Luttinger liquid. Later the model has been used for both the 
normal and superconducting state \cite{3,4,5,6,7,8}, and the usual 
BCS critical 
temperature was re-obtained for the case of a zero anomalous exponent on the 
Green function.

However, the electron-hole channel and the possibility for the occurrence of 
the itinerant electron ferromagnetism was not studied for this model. A recent 
calculation \cite{9} of the magnetic susceptibility of a 1D system, done with 
the renormalization group method and the Monte Carlo simulation, 
showed that the  
magnetic susceptibility $\chi(T)$ becomes constant as $T\ra 0$, but with an 
infinite slope, in contrast with the normal Fermi liquid.

In the present paper we shall assume the existence of a 2D non-Fermi liquid 
described by the Green function:
\be
G(\bk, \o)=\f{g(\a)e^{i\phi}}{\o_c^\a\left[\o-\ve(\bk)+i\d\right]^{1-\a}}
\label{e1}
\ee
where $\a$ is the anomalous exponent, $\phi=-\pi\a/2$ 
is a phase factor introduced 
in order to respect the time reversal symmetry of the Green's function, 
$\ve(\bk)$ is the kinetic energy of 
the electrons and $g(\a)=\pi\a/2\sin{(\pi\a/2)}$ is a factor introduced to 
preserve the equal time anticomutation relations of the electrons. The form 
of the Green's function is assumed to be given by Eq. (\ref{e1}) as long as 
$-\o_c\leq\o\leq\o_c$, $\o_c$ being the energy cutoff specific for the model. 
The exponent $\a$ is non-universal \cite{3,4} and is considered to has a value 
$0<\a<1/2$\cite{1,2,4,7,8}.

In order to study the electron-hole instability we calculate the polarisation 
$\Pi(\bq, \o$ defined by
\be
\Pi(\bq, \o_m)=e^{-i\pi\a}\f{g^2(\a)}{\o_c^{2\a}}\int\f{d^2\bk}{(2\pi)^2}
S(\bk, \bq, i\o_m)
\label{e3}
\ee
where
\bea
S(\bk, \bq, i\o_m)&=&T\sum_n\f{1}{[i\o_n-\xi(\bk)]^{1-\a}}\nonumber\\
&\times&\f{1}{[i\o_n-i\o_m-\xi(\bk-\bq)]^{1-\a}}
\label{e4}
\eea
with $\xi(\bk)=k^2/2m-\mu+nV/2$, $\mu$ being the chemical potential and 
$nV/2$ the Hartree energy of the $n$ electrons. The summation over the 
Matsubara frequencies can be substituted by an complex 
integral and we obtained for the polarisation:

\bea
\Pi(\bq, i\o_m)=2 e^{-i\pi\a}\f{\sin{[\pi(1-\a)]}}
{\pi}\f{g^2(\a)}{\o_c^{2\a}}
\int\f{d^2\bk}{(2\pi)^2}\nonumber\\
\times
\int_{\xi(\bk)}^\infty dx \f{n_F(x)}
{[x-\xi(\bk)]^{1-\a}[x-\xi(\bk-\bq)-i\o_m]^{1-\a}}
\label{e5}
\eea
Following the same way as in Ref. \onlinecite{11} we calculate the 
polarisation as
\bea
\Pi(\bq, i\o_m)&=&2N(0) e^{-i\pi\a}\f{g^2(\a)}{\o_c^{2\a}}\nonumber\\
&\times&
\f{\sin{[\pi(1-\a)]}}{\pi}
\left[\G(2\a-1)S_1(\b, \tilde{\mu})\right.\nonumber\\
&-&i\o_m(\a-1)\G(2\a-2)S_2(\b, \tilde{\mu})\nonumber\\
&-&\
\f{v_F^2q^2}{2}(\a-1)S_3(\b, \tilde{\mu})]
\label{e6}
\eea
where
\be
S_1(\b, \tilde{\mu})=\sum_{m=0}^\infty(-1)^m\f{e^{\b\tilde{\mu}(m+1)}}
{[\b(m+1)]^{2\a}}
\label{e7}
\ee
\be
S_2(\b, \tilde{\mu})=\sum_{m=0}^\infty(-1)^m\f{e^{\b\tilde{\mu}(m+1)}}
{[\b(m+1)]^{2\a-1}}
\label{e8}
\ee
\be
S_3(\b, \tilde{\mu})=\sum_{m=0}^\infty(-1)^m\f{e^{\b\tilde{\mu}(m+1)}}
{[\b(m+1)]^{2\a-2}}
\label{e9}
\ee
with $\tilde{\mu}=\mu-nV/2$.

The paramagnetic susceptibility given by
\be
\chi(T)=-2\mu_0^2 Re\Pi(0, 0)
\label{e10}
\ee
has been calculated as
\be
\chi(T)=-2\mu_0^2N(0)\f{g^2(\a)}{\o_c^{2\a}}\f{2^{2\a}}{\sqrt{\pi}}
\cos{(\pi\a)}\f{\G(\a-1/2)}{\G(1-\a)}S_1(\b, \tilde{\mu})
\ee
where N(0) is the density of states. In the limit $\a\ra 0$ we calculated 
from Eq. (\ref{e7})
\be
S^0_1(\b, \mu)=f_{F-D}(nV/2-\mu)
\label{e12}
\ee
where $f_{F-D}(x)$ is the Fermi-Dirac distribution function, and using this 
result the paramagnetic susceptibility at $T\ra 0$ becomes
\be
\chi_p(T\ra 0)=2\mu_0^2 N(0)
\label{e12bis}
\ee
a result which is in fact the Pauli paramagnetic susceptibility.

The transition temperature can be calculated by considering also the equation 
for the particle density in order to eliminate the effective chemical 
potential $\tilde{\mu}$. The general equation for the electron density is
\be
n=T\sum_n\int\f{d^2\bk}{(2\pi)^2}G(\bk, i\o_n)
\label{e13}
\ee
where $G(\bk, i\o_n)$ is given by Eq. (\ref{e1}). Performing the summation 
over the Matsubara frequencies $\o_n$ and the integral for a constant 
density of states we obtain
\be
n(T, \tilde{\mu})=\f{N(0)g(\a)}{\o_c^\a}\G(\a)\f{\sin{(\pi\a/2)}}{\pi}
S_0(\b, \tilde{\mu})
\label{e14}
\ee
If we consider $S_0(\b, \tilde{\mu})$ well approximated by its value at $\a=0$ 
we have
\be
S_0(\b, \tilde{\mu})\cong \f{1}{\b}\ln{\left[1+e^{\b\tilde{\mu}}\right]}
\label{e15}
\ee
and we get
\be
e^{\b\tilde{\mu}}\cong e^{\b n/B(\a)}-1
\label{e16}
\ee
where 
\bd
B(\a)=\f{N(0)g(\a)}{\o_c^\a}\G(\a)\f{\sin{(\pi\a/2)}}{\pi}
\ed
In order to calculate the critical temperature $T_c$ bellow the itinerant 
ferromagnetism appear we will use the relation
\be
1+V Re\Pi(\bq, i\o_n)=0
\label{e17}
\ee
in the limit $q=0$ and $i\o_n=0$. We mention that according to the 
Mermin-Wagner theorem the ferromagnetic phase is destroyed by the 
fluctuations and $T_c$ is in fact zero. However, the weak three dimensionality 
which is present in High-$T_c$ materials gives us the possibility to use 
Eq. (\ref{e17}) for the calculation of the critical temperature. We can 
introduce this effect phenomenologically taking instead of Eq. (\ref{e17}) 
the equation $1+V Re\Pi(0, 0)=0.01$ which corresponds to $10^2$ times 
enhancement of paramagnetic susceptibility. The qualitative phase diagram is 
not affected by these details. We calculated from Eq. (\ref{e5}) the real 
part of $\Pi(\bq, i\o_n)$ as
\be
Re\Pi(0, 0)=2 N(0)\G(2\a-1) \f{g^2(\a)}{\o_c^{2\a}}\f{\sin{\pi(1-\a)}}{\pi}
S_1(\b, \tilde{\mu})
\label{e18}
\ee
If we consider again that $S_1(\b, \tilde{\mu})$ is well approximated by its 
value for $\a\ra 0$ we have
\be
S_1(\b, \tilde{\mu}=\f{e^{\b\tilde{\mu}}}{e^{\b\tilde{\mu}}+1}
\label{e19}
\ee
which together with Eq. (\ref{e18}) gives for the real part of the 
polarisation the value
\be
Re\Pi(0, 0)=A(\a)\f{e^{\b\tilde{\mu}}}{e^{\b\tilde{\mu}}+1}
\label{e20}
\ee
with
\bd
A(\a)=2\G(2\a-1)\f{g^2(\a)}{\o_c^{2\a}}N(0)\f{\sin{\pi\a}}{\pi}
\ed
Using this result we calculate the critical temperature as
\be
T_c(\a)=-\f{n}{B(\a)}\f{1}{\ln{\left[1-\f{1}{V|A(\a)|}\right]}}
\label{e21}
\ee
From Eq.(\ref{e21}) we get a critical condition for the existence of the 
ferromagnetic order expressed as
\be
V_c<\f{1}{|A(\a)|}
\label{e23}
\ee
In the limit $\a\ra 0$ the critical temperature becomes identical with the one 
obtained for a three dimensional itinerant electron ferromagnet. This result 
is given by the fact that in both cases the integral over the energy variable 
is performed at the Fermi surface using a constant density of states. For a 2D 
electron system a more realistic description will be the one in terms of a 
van Hove density of state. We expect as for the superconducting 
critical temperature calculated 
in Ref. \onlinecite{8} that the ferromagnetic critical temperature will be 
enhanced by the energy dependence of the density of states. Anyway such a 
calculation is much more difficult and the results will be published in 
another work.

\acknowledgements

This work was partly supported by grants from the Instituto Nazionale di 
Fisica della Materia (INFM) under a PRA-HTSC grant (IT), ``Abdus Salam'' 
International Centre for Theoretical Physics under the ``Associate Scheme'' 
(MC). One of us (MC) greatfully acknowledge the hospitality of University of 
Camerino.


\end{document}